\documentclass[journal=nalefd,manuscript=letter]{achemso}
\usepackage{amssymb}
\usepackage{amsmath}
\mciteErrorOnUnknownfalse

\author{Thomas Kanne}
\email{thomas.kanne@nbi.ku.dk}
\affiliation[University of Copenhagen]
{Center for Quantum Devices, Niels Bohr Institute, University of Copenhagen, 2100 Copenhagen, Denmark}

\author{Dags Olsteins}
\affiliation[University of Copenhagen]
{Center for Quantum Devices, Niels Bohr Institute, University of Copenhagen, 2100 Copenhagen, Denmark}

\author{Mikelis Marnauza}
\affiliation[University of Copenhagen]
{Center for Quantum Devices, Niels Bohr Institute, University of Copenhagen, 2100 Copenhagen, Denmark}

\author{Alexandros Vekris}
\affiliation{Sino-Danish College (SDC), University of Chinese Academy of Sciences, China and Center for Quantum Devices, Niels Bohr Institute, University of Copenhagen, 2100 Copenhagen, Denmark}

\author{Juan Carlos Estrada Salda\~na}
\affiliation[University of Copenhagen]
{Center for Quantum Devices, Niels Bohr Institute, University of Copenhagen, 2100 Copenhagen, Denmark}

\author{Sara Lori\`{c}}
\affiliation[University of Copenhagen]
{Center for Quantum Devices, Niels Bohr Institute, University of Copenhagen, 2100 Copenhagen, Denmark}

\author{Rasmus D.\ Schlosser}
\affiliation[University of Copenhagen]
{Center for Quantum Devices, Niels Bohr Institute, University of Copenhagen, 2100 Copenhagen, Denmark}

\author{Daniel Ross}
\affiliation[University of Copenhagen]
{Center for Quantum Devices, Niels Bohr Institute, University of Copenhagen, 2100 Copenhagen, Denmark}

\author{Szabolcs Csonka}
\affiliation[Budapest University of Technology and Economics]
{Department of Physics, Budapest University of Technology and Economics and Nanoelectronics ’Momentum’ Research Group of the Hungarian Academy of Sciences, Budafoki ut 8, 1111 Budapest, Hungary}

\author{Kasper Grove-Rasmussen}
\affiliation[University of Copenhagen]
{Center for Quantum Devices, Niels Bohr Institute, University of Copenhagen, 2100 Copenhagen, Denmark}

\author{Jesper Nygård}
\email{nygard@nbi.ku.dk}
\affiliation[University of Copenhagen]
{Center for Quantum Devices, Niels Bohr Institute, University of Copenhagen, 2100 Copenhagen, Denmark}

\title{Double nanowires for hybrid quantum devices }

\begin{document}

\begin{abstract}
Parallel one-dimensional semiconductor channels connected by a superconducting strip constitute the core platform in several recent quantum device proposals that rely e.g.\ on Andreev processes or topological effects. In order to realize these proposals, the actual material systems must have high crystalline purity and the coupling between the different elements should be controllable in terms of their interfaces and geometry. We present a strategy for synthesizing double InAs nanowires by the vapor-liquid-solid mechanism using III-V molecular beam epitaxy. A superconducting layer is deposited onto nanowires without breaking vacuum, ensuring pristine interfaces between the superconductor and the two semiconductor nanowires. The method allows for a high yield of merged as well as separate parallel nanowires, with full or half-shell superconductor coatings. We demonstrate their utility in complex quantum devices by electron transport measurements.

\end{abstract}


\section{1. Introduction}
The current progress in quantum technology goes hand-in-hand with advances in materials science\cite{giustino20212021}. Within solid state devices, hybrid semiconductor nanowires (NWs) have attracted attention as experimental model systems for quantum computing approaches, including gate-tunable superconducting qubits\cite{aguado2020,Larsen2015,Luthi2018} and devices for studies of Majorana modes that may act as topologically protected qubits~\cite{dassarma2015,aasen2016milestones}. A number of recent proposals are based on multiple coupled nanowires, not yet realized experimentally. For example, pairs of parallel nanowires are at the core of the setup for fractional Majorana fermions that require the one-dimensionality of the individual nanowires combined with inter-wire coupling by crossed Andreev reflections 
\cite{klinovaja2014time}. The resulting “parafermions’’ are expected to enable topological protection on a wider set of operations than Majorana mode qubits thanks to their richer topological structure. Another interesting scenario is the topological Kondo effect \cite{BeriPRL2012,AltlandPRL2013}  that could be realised by joining two NWs in the topological regime by a common superconducting island.
Similar double nanowire geometries could lead to tuning of Majorana regimes \cite{SchradePRB2017}, Majorana box qubits with projection control \cite{plugge2017} and other coupled bound state systems, including Andreev bound state molecules \cite{Yang2014,Scherubl2019}. These proposals all serve to show that building double nanowires (DNWs) could be key for demonstrating a suite of exotic phenomena and testing hypotheses not accessible with only individual superconducting hybrid NWs. 

Harnessing the materials properties of DNWs would be essential since all the proposals require extremely well defined wire geometries and interfaces, including control of the inter-wire spacing, ranging from direct contact and tunneling between the semiconductors to purely electrostatic coupling. Most single nanowire studies have been based on bottom-up synthesized III-V semiconductor nanowires, e.g.\ InAs and InSb, normally grown by the vapor-liquid-solid (VLS) mechanism that yields single-crystal, one-dimensional semiconductors with diameters around 100~nm. By coating these with epitaxial superconducting shells the base materials for the single nanowire qubit schemes were formed \cite{ aguado2020,Krogstrup2015,Lutchyn2018}. However, assembling such nanostructures in pairs or even more complex circuits is a challenge. 
The incentive to grow VLS nanowires in coupled configurations 
has recently led to X-junction and "hashtag'' devices \cite{gazibegovic2017epitaxy,khan2020highly,krizek2017growth} as well as demonstrations of three-dimensional branched "tree'' structures\cite{dick2004synthesis,thornbergAPL2018}, the latter not being compatible with conventional planar device manufacture. 

\begin{figure}
\centering
\includegraphics[width=1\textwidth]{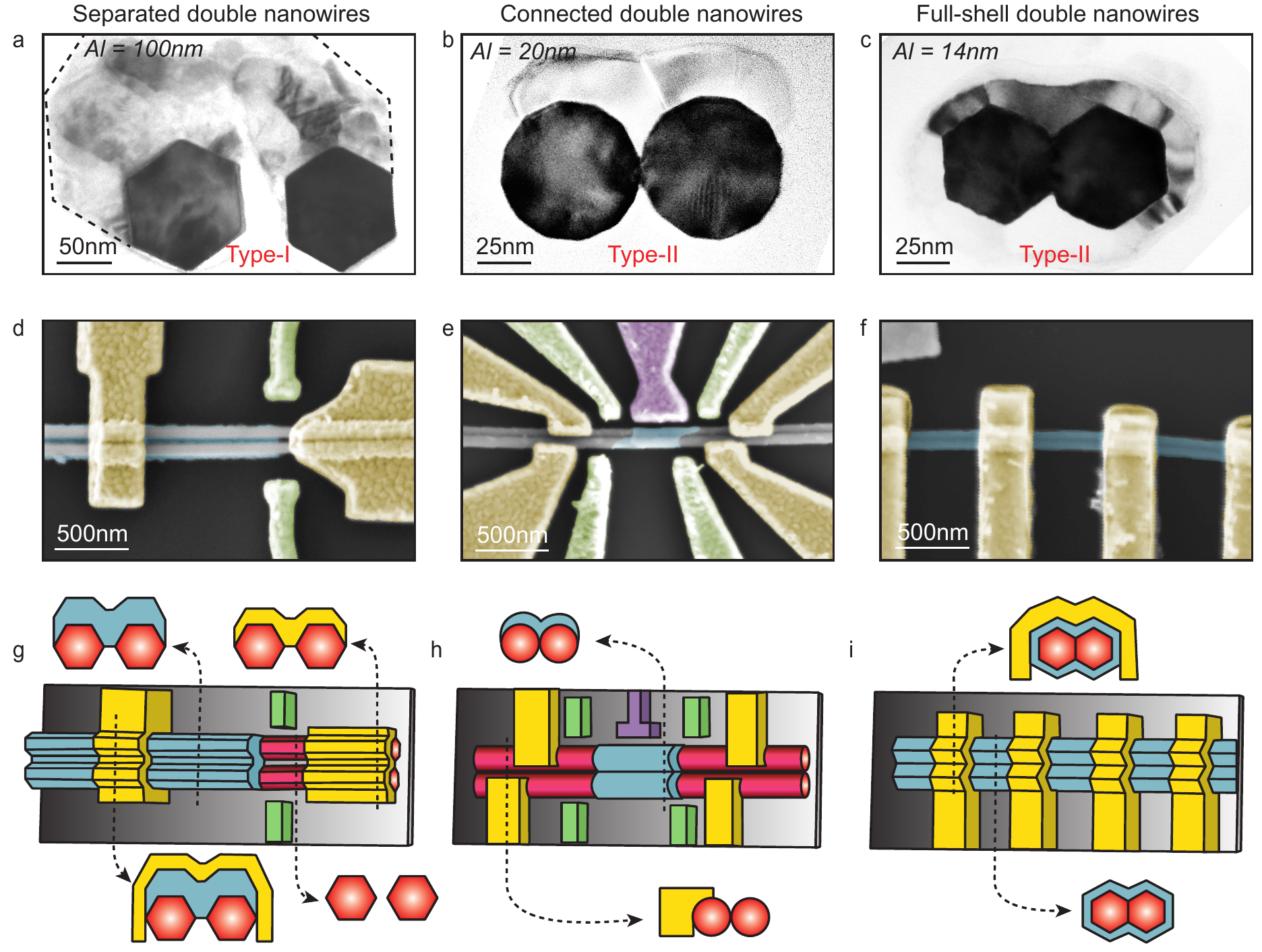}
\caption{\textbf{Double-nanowire configurations and device layouts}. (a-c) Transmission electron micrographs of nanowire cross sections, where pairs of parallel InAs nanowires in three different configurations are coated with an Al film, either on one side (a, b) or around the entire pair as a full shell (c). (d-f) Scanning electron microscopy images of devices; gated normal nanowire-superconductor junction (d), superconducting island with four independent contacts and several gates (e), and full shell double nanowire with a full shell superconducting coating. (f). (g-i) Schematics reflecting the geometry of the devices in panel (d-f); with InAs nanowires (red), aluminum coatings (blue) and gold contacts and gate electrodes (yellow). For (g) the two nanowires are separated (type I) while they are directly connected in h and i (type II). }\label{fig:DNW-fig1} 
\end{figure}

Until now pairs of parallel, identical nanowires coupled by a superconducting layer have not been rationally synthesized. Such hybrid double nanowires (DNWs) would directly fit in standard device fabrication schemes, see Figs.~\ref{fig:DNW-fig1}g-i for examples.
However, so far realization of DNWs has only been possible by randomly depositing NWs on a substrate and localizing pairs that stick together\cite{BabaAPL2015}, or by identifying nanowires that accrete during VLS growth by accident. The first approach has recently been used for devices showing Cooper pair splitting in superconductor-DNW junctions \cite{BabaNJPhys2018,UedaScience2019}, but these techniques do not allow for the crucial controlled in-situ deposition of an epitaxial superconducting shell\cite{aguado2020,Lutchyn2018}. Conversely, in-situ formation of hybrid DNWs would allow for the clean, uniform superconductor-semiconductor interface needed to reach the new regimes discussed above.

In this work we show highly reliable methods for producing DNWs based on pairs of InAs nanowires positioned next to each other during VLS growth. To grow nanowires spaced from each other in a "train track" geometry (denoted type I), the growth parameters are set to yield rigid nanowires, ensuring that the separation between the nanowires remains constant from top to bottom. 
These structures are realized with a superconductor (Al) half-shell and full-shell configuration while still retaining a separation between the nanowires; see a cross section in Fig.~\ref{fig:DNW-fig1}a. In a different approach (denoted type II), the growth parameters are set to yield longer and comparably less rigid nanowires. Here the nanowire pairs connect in their upper segments before the metal deposition as van der Waals forces clamp the two nanowires together in an "Eiffel tower" configuration, yielding directly connected NWs (Fig.~\ref{fig:DNW-fig1}b-c). The DNW geometry can be easily tuned by change of growth parameters and adjustment of the initial Au particle dimensions and inter-particle separation. 


Below we describe in more detail the synthesis and design criteria behind these structures and demonstrate that they can be readily implemented in selected quantum devices (Figs.~\ref{fig:DNW-fig1}d-f). We note that these materials have recently been exploited for studies of Andreev molecular states \cite{Kurtossypreprint2021}, double quantum dot Josephson junctions \cite{Vekrispreprint2021} (Fig.~\ref{fig:DNW-fig1}d) and Little-Parks oscillations using full-shell DNWs \cite{VekrispreprintLP2021} (Fig.~\ref{fig:DNW-fig1}f), demonstrating that ready-made DNWs are useful in functional quantum devices. 
Finally, we show designs that extend beyond the double-wire configurations and may enable experiments that can by no means be realized by 
serendipitous DNW formation.

\section{2. Growth of parallel nanowires}

Molecular beam epitaxy (MBE) is used to grow double InAs nanowires with a Wurtzite crystal structure along the [0001]B direction on InAs (111)B substrates. The double nanowires are grown via the vapor-liquid solid (VLS) mechanism and catalyzed by a pair of electron-beam lithography (EBL) defined Au particles. To facilitate growth of pairs of nanowires in very close proximity, the Au particles are engineered and patterned according to the intended inter-nanowire geometry (type I/II), while the MBE growth procedure itself is similar to one used for conventional single nanowires.
To have precise control of the Au catalyst particles (disc radius $r_D$) and their separation ($D_{Au}$), multiple electron beam dot exposures ("single shots") are positioned in a pre-defined pattern with a dot separation of 20~nm. The typical disc thickness and radius is around 20~nm and 25-50~nm, respectively (see Supplementary Information for details on Au particle formation and growth procedures). 
Growth of double nanowires follows largely the single nanowire growth dynamics, however, the close proximity of the nanowires reduces the contribution of adatoms to the liquid due to the shared adatom collection area on the substrate \cite{madsen2013experimental}. In InAs nanowire growth, the main adatom contribution consists of In atoms and thus a higher effective V/III flux ratio is attained for double nanowires. With the current growth parameters this causes the double nanowires to be thinner compared to single nanowires. This effect can readily be counteracted in MBE by change of nominal fluxes. 
Growth parameters and a comparative analysis of single and double nanowire growth are presented in Supplementary Information, section 4. 
After VLS growth of the semiconductor nanowires, an aluminum layer is deposited onto nanowires by electron beam evaporation in a separate metal deposition chamber connected directly to the MBE system. The evaporation of the aluminum film follows the procedure outlined in Ref. \cite{Kanne2020} where a low substrate temperature ($T_{sub} \sim 120$~K) and high rate (3~Å/s) are found to ensure a pristine bi-crystal interface while maintaining a flat and continuous morphology\cite{Krogstrup2015,Kanne2020}.


\begin{figure}
\centering
\includegraphics[width=1\textwidth]{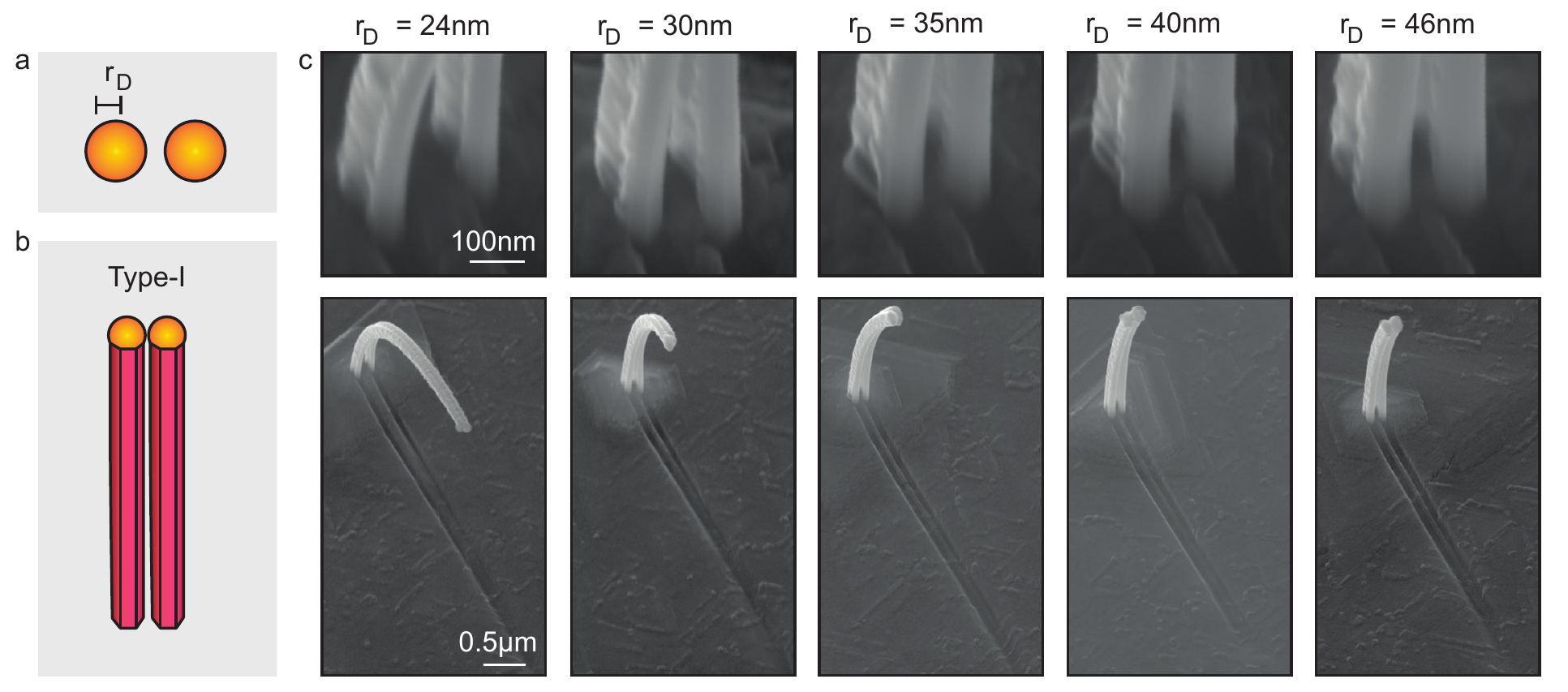}
\caption{{\bf InAs nanowire pairs grown utilizing the type I approach.} (a) Initial Au particle configuration. (b) Schematic of double nanowire with a constant inter-nanowire separation. (c) SEM micrographs of nanowires grown with a center to center Au particle separation of $D_{Au}=160$~nm and different Au disc radii $r_\mathrm{D}$. Top-row micrographs show the bottom part of the nanowire pairs and the lower row shows the full nanowire system and shadows cast on the substrate during metal evaporation.  }\label{fig:DNW-fig2}
\end{figure}

Arrays of double nanowires are defined by the lithographic patterning of Au particles prior to the growth. In order to obtain various double nanowire configurations we define fields where the separation between the centers of the Au particles $D_{Au}$ is varied from, e.g., 50~nm to 300~nm at intervals of 10~nm while also varying the Au disc radius. 
With this strategy, we find nanowires that are too far apart and therefore stand as single nanowires as well as Au particles that are too close and combine to nucleate a single nanowire. However, importantly, we ensure the formation of extended arrays of parallel nanowires with appropriate inter-nanowire spacings, independent of fluctuations between growths and minor adjustments of growth parameters. 

Figure \ref{fig:DNW-fig2} shows InAs nanowires grown with varied Au particle radii and a constant initial Au-droplet center-center separation of 160~nm. To close the gap between the nanowire pairs, a thick layer ($\sim 100$~nm) of Al was deposited with an angle of $\sim 20^{\circ}$ with respect to the normal of the gap between the nanowires. The SEM micrographs in Fig.~\ref{fig:DNW-fig2}c are obtained at an angle opposite to the Al deposition direction in order to highlight the morphology of the nanowires. The top row focuses on the base of the nanowires, while the bottom row reveals the shading effects on the substrate. From Fig.\ \ref{fig:DNW-fig2}, pronounced thickness-dependent bending of the nanowires is observed as seen also for single nanowires\cite{Krogstrup2015,Bjergfelt2019c}. By examining the metal deposited on the substrate, it appears that all semiconductor nanowires grow separately and in a later stage connect via the Al film, in accordance with TEM studies of cross-sections such as Fig.~\ref{fig:DNW-fig1}a. For the specific conditions used for the growths in Fig.\ \ref{fig:DNW-fig2}, diameters greater than 80 nm yield a uniform nanowire separation, i.e.\ type I double nanowires.
We also observe from Fig.~\ref{fig:DNW-fig2}c that longer nanowires with smaller diameters
 merge near the top. These flexible nanowires\cite{Erdelyi2012} are accidentally clamped by van der Waals forces when vibrating in the growth chamber.

\begin{figure}
\centering
\includegraphics[width=1\textwidth]{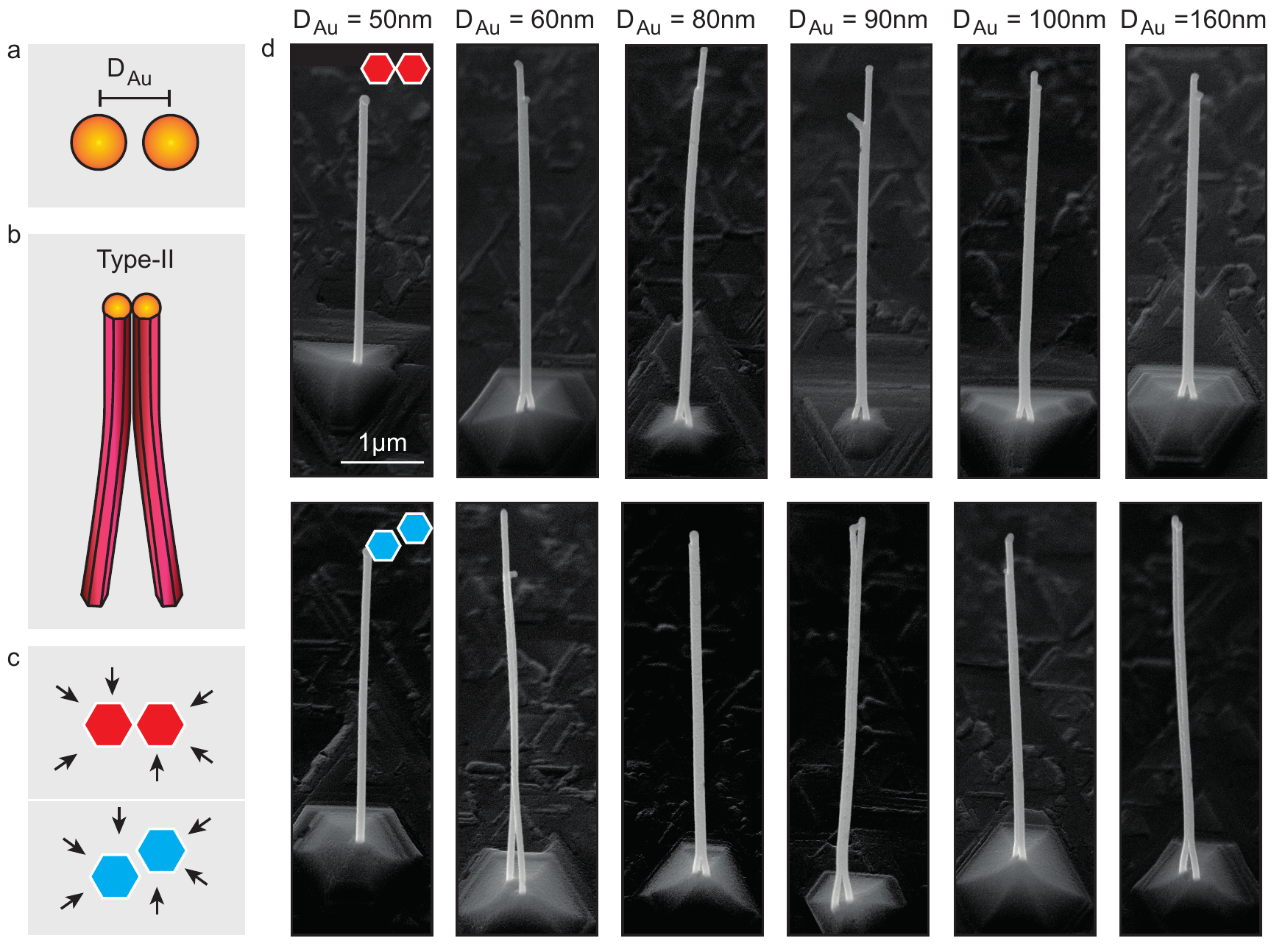}
\caption{{\bf Parallel InAs nanowires utilizing the type II growth approach with a full shell of Al.} (a) The initial Au particle disk separation ($D_{Au}$). (b) The type II merging scheme. (c) Top-view schematic of a corner-to-corner (red) and a facet-to-facet (blue) configurations with a $\sim $~27~nm full Al shell resulting from six depositions perpendicular to nanowire facets. (d) SEM micrographs of nanowire pairs in the corner-to-corner and facet-to-facet configurations for Au particles formed by one single electron beam exposure and a center to center particle separation according to the text above the micrographs. The scales are all the same and noted in the first micrograph.}\label{fig:DNW-fig3}
\end{figure}

To further enhance this behaviour and thus facilitate predominantly type II merged nanowires, the growth parameters are adjusted to form thinner and longer nanowires. To decrease the average nanowire diameter for the growth shown in Fig.~\ref{fig:DNW-fig3}d, we increased the V/III flux ratio ($\sim$ 20) by increasing the As$_4$ flux during the axial growth step (see Supplementary Information section 4 for details). 
This yielded nanowires with diameters of $\sim 50$~nm for Au droplets with $r_{D} \sim 24$~nm and $\sim 130$~nm for $r_{D} \sim 100$~nm. To fully cover the nanowires by Al, six $\sim 5$~nm Al layers were deposited with 60$^{\circ}$ rotations between each evaporation, as schematically shown in Fig.~\ref{fig:DNW-fig3}c. Single directional evaporations are equally feasibly as shown later in Fig.\ \ref{fig:DNW-Fig5}.
In Fig. \ref{fig:DNW-fig3}d SEM micrographs of nanowires are shown for facet-to-facet and corner-to-corner configurations (see blue and red schematics, respectively, in Fig.~\ref{fig:DNW-fig3}c) with  center-to-center Au particle separation varied from 50 to 160~nm and $r_{D}=24$~nm. The initial Au particles are positioned according to the underlying crystallographic basis of the substrate to achieve the different configurations.   
For all separations above 50~nm the images show consistent parallel nanowires in both configurations. 
A TEM micrograph of a full-shell type II cross section is shown in Fig. \ref{fig:DNW-fig1}c. 

Occasionally, inter-nanowire spacings and nanowire dimensions are observed to differ from the intended range, and some Au particles are found missing in post-growth SEM characterization. In rare cases the nanowires do not grow along the intended [0001]B direction. With these aberrations 
we find a yield of 50-60~\% for type II merged nanowires as documented in Supplementary Information section 3. For these growths hundreds of identical double nanowires are found in numerous arrays across the growth substrates. 
We believe that most of the mishaps can be related to variations in the Au nanoparticle lithography process. The present work is not focused on optimization of the large scale yield, however, based on the demonstrated results we expect that near unity yield for double nanowire growth can be reached with optimization of the substrate preparation.

The growth dynamics for paired InAs nanowires practically follows that of conventional single nanowires, thus modifications of growth parameters can readily be used to control the crystal quality and nanowire dimensions, which is important for fulfilling the requirements in the proposed device designs. We find that it is easier to achieve a high yield of type II merged nanowires compared to type I. 
For the latter type, it is a challenge to define Au particles that can catalyze large (thick) nanowires whose rigidity prevents type II merging and yet are spaced close enough that the subsequent metal deposition can fill the gap between the vertical wires.  One can overcome this by radially overgrowing nanowires that are nucleated by Au particles with a large height/diameter aspect ratio. Such particles show smaller displacement when they acquire the shape of a spherical cap at the growth temperatures. This approach is demonstrated in the Supplementary Information section 2.

\section{3. Electrical transport characterization}

\begin{figure}
\centering
\includegraphics[width=0.4\textwidth]{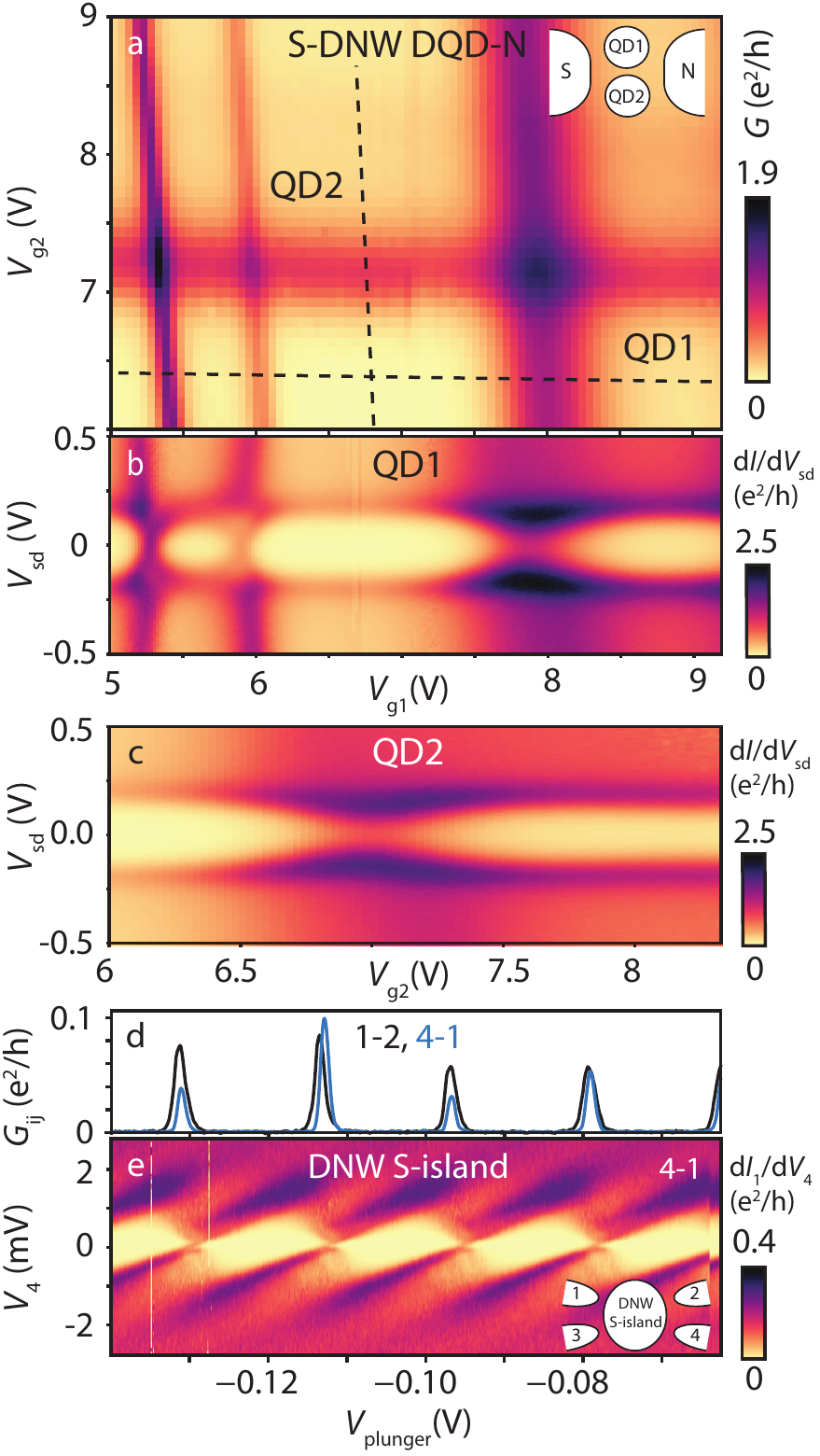}
\caption{\label{fig:DNW-fig4}\small \textbf{Electrical transport measurements of two hybrid double nanowire (DNW) devices.} (a) Linear conductance $G$ versus side gate voltages $V_{\mathrm{g1}}$ and $V_{\mathrm{g2}}$ of a superconductor (S)-parallel double quantum dot-normal metal (N) device (see inset and Fig.~\ref{fig:DNW-fig1}(d,g) for device geometry, in this case with merged normal leads). The nearly horizontal and vertical conductance lines reveal the transport resonance conditions for the two quantum dots controlled by the respective gates. (b-c) Bias spectroscopy plots of differential conductance 
${\rm d}I/{\rm d}V_{\rm sd}$ versus bias 
$V_{\rm sd}$ and gate $V_g$ of quantum dot 1 (2) (QD1, QD2) along the horizontal (vertical) dashed lines in (a). The maps show that both quantum dots are coupled to a common superconducting lead (gap $\Delta \sim 0.2$~meV).  
A small  gate shift has occurred between measurement (a) and (b,c). (d) Conductance $G$ versus plunger gate $V_{\rm plunger}$ of a superconducting island (See inset in (e) and Fig.~\ref{fig:DNW-fig1}(e,h) for device design). The plot shows Coulomb blockade behavior for two pairs of electrodes, i.e.\ upper nanowire (1-2) and interwire (4-1) nanowire transport. 
(e) Bias spectroscopy showing Coulomb blockade diamonds related to transport through the S-island via leads 4-1. All measurements were performed at 30\,mK. }
\end{figure}
We now turn to electrical transport characterization of the hybrid double nanowire system, focusing here on low temperature transport measurements of two different types of devices in the quantum regime. 
The objective is to demonstrate that decoupled double nanowires and multi-terminal device functionalities can be achieved (more advanced functionalities are demonstrated in forthcoming dedicated publications \cite{Kurtossypreprint2021, Vekrispreprint2021,VekrispreprintLP2021}). Prior transport experiments on parallel nanowires have been based on individual nanowires that were accidentally merged during deposition \cite{BabaNJPhys2018,UedaScience2019}.
In contrast, we transferred ready-made DNWs from the growth substrate to pre-patterned device substrates using a micromanipulator under an optical microscope. 

The first example is a parallel double quantum dot device formed between the superconducting in-situ deposited 100\,nm Al and an ex-situ evaporated normal metal (Ti/Au) covering both wires (Fig.~\ref{fig:DNW-fig1}d). Prior to deposition of the normal metal, the Al on part of the nanowires has been partly removed by etching (Al etchant Transene D). The inset of Fig.~\ref{fig:DNW-fig4}a shows a conceptual schematic of the resulting device, where side and back gate voltage are appropriately tuned to make a superconductor (S)-parallel double quantum dot (DQD)-normal (N) device (see also Fig.~\ref{fig:DNW-fig1}g device schematics), however, in this case with the normal leads merged)\cite{BabaAPL2015,DeaconNatComm2015}. The conductance versus side gate voltage map plotted in Fig.~\ref{fig:DNW-fig4}a shows two slopes of conductance resonances (nearly horizontal/vertical), consistent with transport through two parallel quantum dots. To address the superconducting properties of the in-situ deposited Al segment, we perform bias spectroscopy along the dashed traces in Fig.~\ref{fig:DNW-fig4}a to independently tune the occupation of each quantum dot. For both QDs, we observe clear signs of tunneling via the superconducting coherence peaks at $eV_\mathrm{sd} \sim \pm 200$\,$\mu$V shown in Fig.\ \ref{fig:DNW-fig4}b-c in agreement with the expected superconducting energy gap of an Al film. Importantly, the two QDs appear only weakly coupled, with an upper bound given by the width of the resonances (Fig.~\ref{fig:DNW-fig4}a)\cite{vanderWielRevModPhys2002}, in accordance with the separated ''train track" nanowire geometry (type I). Moreover, the behavior in particular for QD1 seems to reflect Yu-Shiba-Rusinov states in different coupling regimes\cite{DeaconPRL2009,LeeNatNano2014,JellinggaardPRB2016}. 
The superconductor-DQD-normal metal device geometry is easily extended to S-parallel DQD-S Josephson junctions, where supercurrent and bound states behavior can be addressed \cite{Vekrispreprint2021,Kurtossypreprint2021}.

The second example is a superconducting island defined across the two nanowires, a geometry relevant for several device proposals in the topological and non-topological regimes \cite{BeriPRL2012,AltlandPRL2013,GalpinPRB2014,PapajPRB2019}.
The SEM picture in Fig.\ \ref{fig:DNW-fig1}d shows a ''Tetris" island geometry (potentially relevant to decrease the overlap of Majorana end states on the island if it were in the topological regime), while the data discussed below were obtained on a rectangular island of length 330\,nm. 
The superconducting island is defined by lithographic patterning and an etching step of the 17~nm thick Al film. Subsequently normal metal electrodes are defined on each nanowire following standard EBL and metal deposition recipes. The device schematics is shown in Fig.\ \ref{fig:DNW-fig4}e, where the superconducting island --- in contrast to a single nanowire device --- has contact to four leads. In Fig.\ \ref{fig:DNW-fig4}d we show the two-terminal linear conductance of two combinations of leads (1-2, 4-1) versus side gate voltage with the other terminals floating. It clearly displays Coulomb blockade behavior in the weakly coupled regime. This phenomenon is further confirmed by the bias spectroscopy measurements shown in Fig.\ \ref{fig:DNW-fig4}e for one of the combinations revealing Coulomb blockade diamonds. 
The superconducting properties cannot be deduced from this data alone, but additional measurements in a different coupling regime reveal the superconducting properties in the even-odd regime via finite bias negative differential conductance features \cite{higginbotham2015parity} facing the odd occupation of the island (to be reported in a separate work). The demonstration of Coulomb blockade via different leads to the same superconducting island is promising for investigations of the topological Kondo effect, Majorana box qubits\cite{plugge2017} and specific proposals based on the hybrid double nanowire geometry \cite{BeriPRL2012,AltlandPRL2013,GalpinPRB2014,GaidamauskasPRL2014,KlinovajaPRB2014,EbisuProg2016, ReegPRB2017,SchradePRB2017,ThakurathiPRB2018, SchradePRL2018,PapajPRB2019, DmytrukPRB2019,KotetesPRL2019,HaimPhysRep2019,ThakurathiPRR2020}. Yet other experiments addressing the transport properties of the same double nanowire materials 
demonstrate coupled Andreev states\cite{Kurtossypreprint2021} and Yu-Shiba-Rusinov physics\cite{Vekrispreprint2021} emerging in the double nanowire Josephson junction geometry. Furthermore, the full-shell double nanowire devices such as the one represented by Figs.\ \ref{fig:DNW-fig1}c,f 
show Little-Parks oscillations in low-temperature transport measurements under a parallel magnetic field\cite{VekrispreprintLP2021}, underlining that a broad range of quantum transport phenomena can be addressed in these parallel nanowire systems.


\section{4. Beyond double nanowires}
\begin{figure}
\centering
\includegraphics[width=0.9\textwidth]{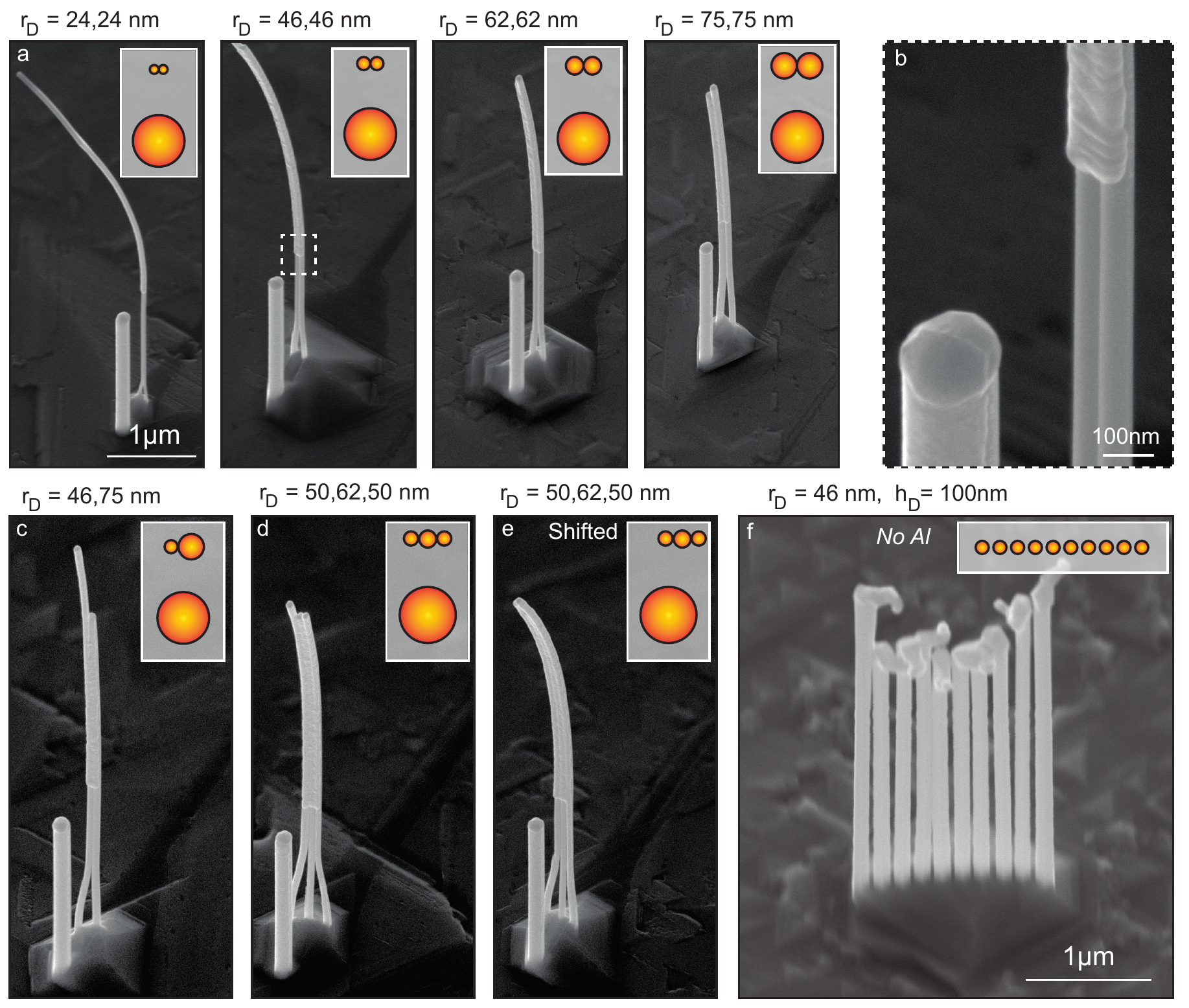}
\caption{\label{fig:DNW-Fig5}\textbf{ 
SEM micrographs of parallel nanowires in complex configurations.} All nanowires were grown on the same wafer and all with $\sim$ 20nm Al (except panel f). (a) In-situ shadowed parallel nanowires with varied dimensionality formed by different Au partices sizes. (b) Zoom on the second panel in (a). 
The sharp shadow junction is visible as well as the rounded nanowires obtained by tuning the growth parameters. 
(c) Parallel nanowire pair with different diameters. 
(d) Three nanowires merged around the center nanowire that has an increasing diameter. (e) Three nanowire bundle similar to (d), however shifted to ensure that one nanowire is not shadowed. (f) 
Linear array of 10 parallel nanowires without metal coating.
}
\end{figure}

Growth substrate preparation by EBL allows the controlled fabrication of several different nanowire structures as shown in Fig.~\ref{fig:DNW-Fig5}. Here in-situ nanowire shadowing of the subsequently deposited superconductor was furthermore implemented to allow for fabrication of hybrid devices without metal etching steps\cite{Carrad2019}. By fine tuning catalyst size and positioning via the EBL step it is possible to obtain a range of nanowire dimensions (Fig. \ref{fig:DNW-Fig5}a) and also to realize more complex heterostructures such as the shadowed triplet structure seen in Fig. \ref{fig:DNW-Fig5}d-e. In Fig. \ref{fig:DNW-Fig5}c we show that variations of the Au particle volume may be used to form nanowires with different dimensions and with that determine the overall morphology of the pairs. In turn this can be a way to provide different transport characteristics of the two nanowires, e.g.\ different numbers of one-dimensional subbands in DNWs with two different diameters. In Fig. \ref{fig:DNW-Fig5}e the triplet structure is shifted slightly to one side so that one of the three nanowires is fully coated and the remaining two are half shadowed. Finally, Fig. \ref{fig:DNW-Fig5}f showcases the feasibility of more complex structures, here exemplified by 10 nanowires in a row.


\section{Conclusions}
In conclusion, the required DNW-superconductor hybrids have been successfully developed and can be grown with a high yield. The electrical transport has elucidated the inter-wire couplings and the properties of the DNW-superconductor hybrids. Focus has so far been on realizing Al/InAs-based DNWs by MBE, but the concepts can be transferred to other III-V methods and materials that maintain an epitaxial relationship between the nanowires and the growth substrate, such as e.g.\ InSb and InP wires as well as other growth systems that are also utilizing the VLS mechanism for nanowire growth, e.g.\ metalorganic vapour phase epitaxy or chemical beam epitaxy. Likewise, the in-situ approach used here will also allow for several different metal (superconductor) coatings 
and other in-situ depositions, e.g.\ of dielectrics or magnetic materials. 
Finally, DNWs can also be formed from NWs with axial heterostructures.

The parallel nanowire configurations demonstrated in this work matches the requirements for several proposals within hybrid superconductor-semiconductor devices, in particular experiments within the topological regime that cannot be performed with single hybrid nanowires.  
 Moreover, DNWs could enable experiments on Coulomb drag induced supercurrents\cite{duan1993} in a system with low screening. Finally, bare multiwire arrangements with several parallel wires, based on structures shown in Fig.~\ref{fig:DNW-Fig5}f, mimic the setup proposed for studying the fractional quantum Hall effect in a quantum wire array \cite{KanePRL2002}.

\section{Author contributions}
CS, KGR, JN conceived the in-situ DNW concept, TK, DO, MM, RDS, DR, JN developed growth schemes and performed structural characterizations, AV, SL, JCES, KGR, JN performed transport measurements, JN supervised the project, TK, KGR, JN wrote the manuscript with input from all authors.

\begin{acknowledgement}
This work was funded by the European Union’s Horizon 2020 research and innovation programme QuantERA project no.~127900 (SuperTOP) and FETOpen grant no.~828948 (AndQC), the Carlsberg Foundation, the Niels Bohr Institute, the Villum Foundation project no.~25310 and the Ministry of Innovation and Technology and the NKFIH
within the Quantum Information National Laboratory
of Hungary and by the Quantum Technology National
Excellence Program (Project Nr. 2017-1.2.1-NKP-2017-
00001), NKP-20-5 New National Excellence Program. The Center for Quantum Devices is supported by the Danish National Research Foundation. J.C.E.S. acknowledges funding from the European Union’s Horizon 2020 research and innovation program under the Marie Sklodowska-Curie grant agreement No. 832645.
The authors thank M.~Aagesen, M.~Burello, I.~Nielsen, J.~Paaske, G.~Steffensen, M. Wauters, C.B.~S\o rensen, D.~Laroche, D.~Kjær, O.~Kurtossy, Z.~Scherubl, C.~Schrade, J.~Sestoft, and P.~Makk for assistance and discussions.
\end{acknowledgement}

\bibliography{TeorirefsDNW.bib}
\begin{suppinfo}
Supplementary data, micrographs and discussions of parallel nanowire synthesis are found in the Supplementary Information document.



\end{suppinfo}
\end{document}